\documentclass[aps,showpacs]{revtex4} 
\usepackage{amssymb}
\usepackage{graphicx}
\usepackage{dcolumn}
\usepackage{amsmath}
\begin{document}
\title{Electronic properties of the ternary system YbPd$_2 $Ge$_{2-x}$
}
\author{\fbox{V.N. Nikiforov}$^1$, V.V. Pryadun$^1$, A.V. Gribanov$^1$, V.Yu. Irkhin$^2$}
\email{Valentin.Irkhin@imp.uran.ru}
\affiliation{
$^1$Moscow State University, 119899 Moscow, Russia}

\affiliation{$^{2}$M. N. Mikheev Institute of Metal Physics, 620108 Ekaterinburg, Russia}

\begin{abstract}
The data on the electronic and magnetic properties of the ternary ytterbium-based alloy  YbPd$_2 $Ge$_{1.7}$ are presented and compared with those for YbPd$_2 $Ge$_{2}$ and YbPd$_2 $Ge compounds. YbPd$_2 $Ge$_{1.7}$ demonstrates a moderate heavy-fermion behavior with linear specific heat coefficient $\gamma \simeq $ 100 mJ/mol K$^2$. Magnetic susceptibility is large and demonstrates nearly Curie behavior. Resistivity shows a power-law T$^{1.2}$-dependence  at not too low temperatures $T$. Thermoelectric power has non-monotonous behavior and changes its sign at low $T$.
\end{abstract}
\maketitle

\section{Introduction}
After investigation of unconventional electronic properties of classical heavy-fermion and Kondo-lattice systems during 1980-s, ternary compounds are extensively studied with especial attention to   magnetism \cite{I17}. Of particular interest are intermetallic ytterbium-based systems which demonstrate a number of peculiarities.
In some ytterbium compounds, manifestations of the Kondo effect coexist with ferro- and antiferromagnetic ordering.
In a number of ytterbium systems nearly temperature-linear resistivity is observed, indicating non-Fermi-liquid (NFL) behavior.
It is interesting that this feature of resistivity is observed in the system YbMn$_2 $Sb$ _2 $ with nonmagnetic ytterbium, where normal Kondo effect should be absent. These anomalies may be associated with scattering on pseudospin degrees of freedom \cite{NikiforovYb}.

Nikiforov et al \cite{NikiforovFTT,NikiforovFTT1} investigated the ytterbium-based ternary compounds YbPd$ _2 $Ge, YbPd$ _2 $Si, YbPdGe, YbPdSi, YbPtGe, synthesized for the first time, and also previously known compounds YbPd$ _2 $Si$ _2 $ and YbPd$ _2 $Ge$ _2 $, the latter one demonstrating superconductivity  \cite{Hull}. From galvanomagnetic properties and magnetic susceptibility it was found that germanium compounds YbPdGe, YbPtGe and YbPd$_2$Ge exhibit ferromagnetic ordering at low temperatures (near 16, 10 and 12 K, respectively).
Later magnetic and thermodynamic properties were studied thoroughly. In Ref. \cite{YbPtGe} the values of the Curie temperature $ T_C = 5.4 $ K, and linear specific heat coefficient $ \gamma = $ 209 mJ/mol K$^2 $ below $ T_C$ were found for YbPtGe. The magnetic entropy at $ T_C $ is strongly reduced (which is characteristic for Kondo lattices)  and makes up $\mathcal{S}=$ 0.52 $ R \ln 2 $ ($R$ is the universal gas constant). For YbPdGe, the values $ T_C $ = 11.4 K, $\mathcal{S}=$ 0.7 $ R \ln 2 $  and  $ \gamma $ = 150 mJ/mol K$^2 $ were obtained \cite{YbPdGe1}. The situation in these systems resembles CeRuSi$_2 $ where ferromagnetic ordering coexists with the Kondo effect too \cite{CeRuSi2}.

The paper \cite{YbPdGe} discussed various Kondo manifestations in YbPdGe and YbPtGe, including logarithmic contribution to the resistivity, and a comparison of the transport properties with the corresponding cerium systems CePdGe and CePtGe (formally trivalent ions of cerium, and ytterbium are similar because correspond to electrons or holes in 4f-shell).
As shown in Refs. \cite{CeTX}, CeTX compound (T = Pd and Pt, X = Ga, Ge and Sn) are antiferromagnetic Kondo systems at low temperatures,  in contrast with YbTGe.
In the review paper \cite{Flouquet}, a difference between these cases is discussed in detail for the example of YbRh$ _2 $Si$ _2 $ which is a  typical antiferromagnetic heavy fermion ytterbium compound. Despite similarities with CeRh$ _2 $Si$_2 $ (similar temperatures Kondo $T_K $ and resistivity to high amplitude, $T$-linear resistivity above the Neel temperature $ T_N $), there are notable differences, which can be related to more localized features (smaller hybridization with conduction electrons), as well as to a strong spin-orbit interaction in the case of 4f-orbitals of ytterbium.

YbPdSi is a ferromagnetic heavy-fermion system with $ T_C = 8 $ K and $ \gamma \simeq$ 300 mJ/mol K$^2 $ \cite{YbPdSi}.
Studying YbTGe compounds (T = Rh, Cu, Ag) \cite{YbTGe} showed that YbRhGe is an antiferromagnet with $ T_N = 7 $~K, YbCuGe is a ferromagnet with the moment $ 0.7 \mu_B $ and $ T_C = 8 $~K; YbAgGe was found to have a very large $ \gamma = $ 570 mJ/mol K$^2 $. In YbAgGe, canted low-temperature magnetic ordering was found \cite{YbAgGe}.
According to \cite{YbAgGe1}, YbAgGe is a frustrated heavy fermion antiferromagnet with a complex magnetic phase diagram and NFL behavior.

In the present paper we discuss the electronic and magnetic properties of the ternary ytterbium-based alloy  YbPd$_2 $Ge$_{1.7}$.

\section{Experimental}

The sample had the mass $m=16.3$ mg and the size of 0.8x0.7x1.0 mm$^3$.
The temperature dependence of magnetic moment of the sample was measured by a SQUID magnetometer of Quantum Design MPMS-5.

The  measurements of transport properties were carried out over temperature range 1.5--390 K  by the Quantum Design Physical Property Measurement System. This approach provided simultaneous
measurements of  electrical resistivity, thermal conductivity and Seebeck coefficient (thermopower at zero electric current)  by monitoring both the temperature and voltage drop across a sample as a heat pulse is applied to one end.
Electrical resistance  was measured employing standard four-contact geometry at a constant current. Measurements were performed under high vacuum (about 10$^{-4}$  torr) using a four-probe lead configuration.
The Seebeck coefficient  was measured when they were disconnected from the current source. The heat flow in the samples was produced by an electrical heater. The thermally induced voltage was measured through the same potential contacts used in measuring resistance.
The temperature drop along the sample was measured using the differential manganin-constantan
thermocouple.


The thermal conductivity was measured by the method of the longitudinal stable state. The
temperature difference across the samples was produced by the electrical heater. The sample was
clamped to the other end of the copper base. Heat losses from the sample were minimized by
evacuating the chamber.
The temperature of the copper base was obtained using a chromel–constantan thermocouple. The temperature difference across the samples was measured by the differential manganin–constantan thermocouple. Since both manganin and constantan have low values of thermal conductivity,
 this thermocouple did not disturb significantly the heat flow pattern in the sample.

\section{Results for electronic and magnetic properties}

Magnetic moment of the sample $M$ at $T=2$ K in the field of 9000 G  is practically saturated and makes up about 0.02 emu. Magnetic susceptibility   demonstrates nearly Curie behavior (Fig.1). The effective  magnetic moment per Yb atom $\mu_{\rm eff}$   determined
from the fit  $M=N H \mu^2_{\rm eff}/3 k_B T$ ($H$ is magnetic field, $N$ is the number of atoms, $ k_B$ is the Boltzmann constant)  makes up about 1$\mu_B$.

The  low-temperature  $T$-linear electronic specific heat was obtained by the standard  extrapolation starting from temperatures above 20 K to low temperatures, as shown in Fig. \ref{gamma}. The coefficient  $\gamma \simeq $ 100 mJ/mol K$^2$  is  markedly enhanced in comparison with standard metals. 

\begin{figure}[htbp]
\includegraphics[width=3.3in, angle=0]{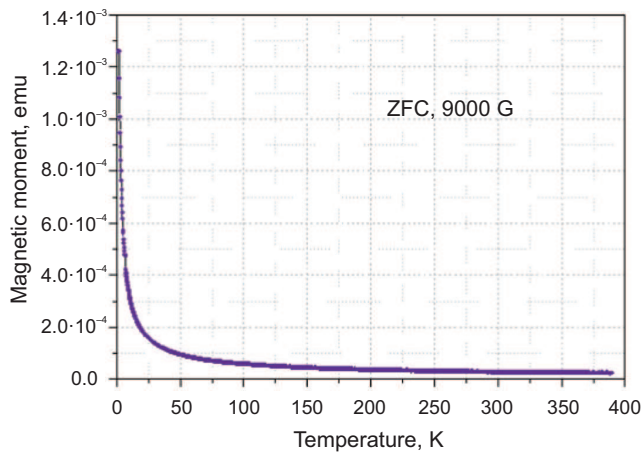}
\caption{The temperature dependence of magnetic moment of the sample}
\label{m}
\end{figure}

\begin{figure}[htbp]
\includegraphics[width=3.3in, angle=0]{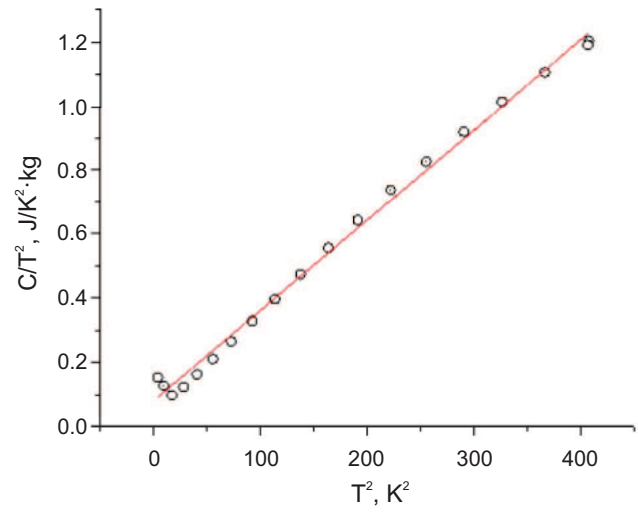}
\caption{The determination of $T$-linear electronic specific heat}
\label{gamma}
\end{figure}

\begin{figure}[htbp]
\includegraphics[width=3.3in, angle=0]{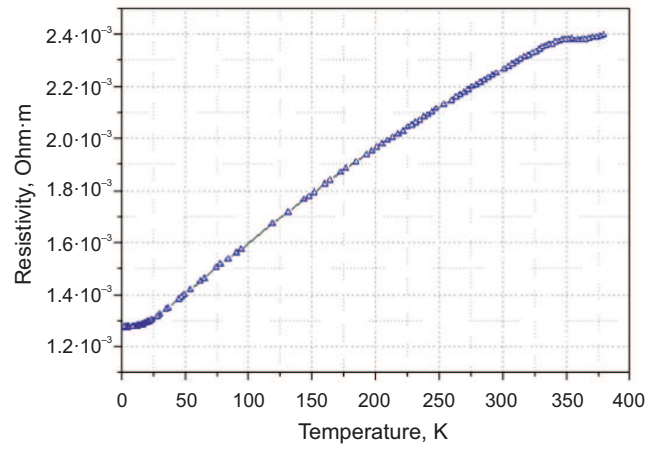}
\caption{The temperature dependence of  resistivity in a wide temperature region }
\label{resist}
\end{figure}

\begin{figure}[htbp]
\includegraphics[width=3.3in, angle=0]{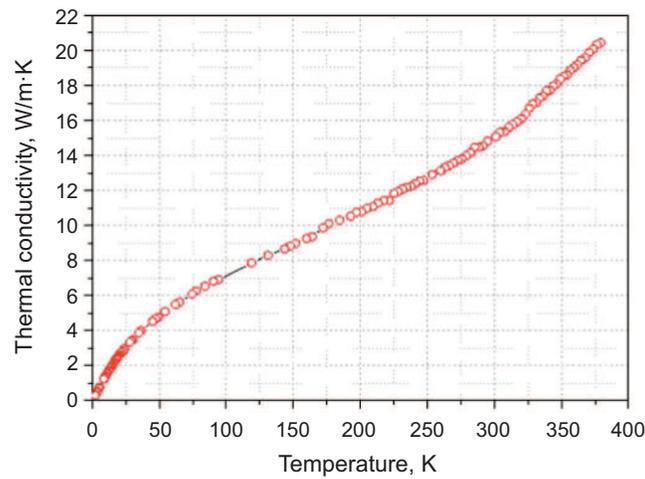}
\caption{The temperature dependence of  thermal conductivity  }
\label{therm}
\end{figure}

\begin{figure}[htbp]
\includegraphics[width=3.3in, angle=0]{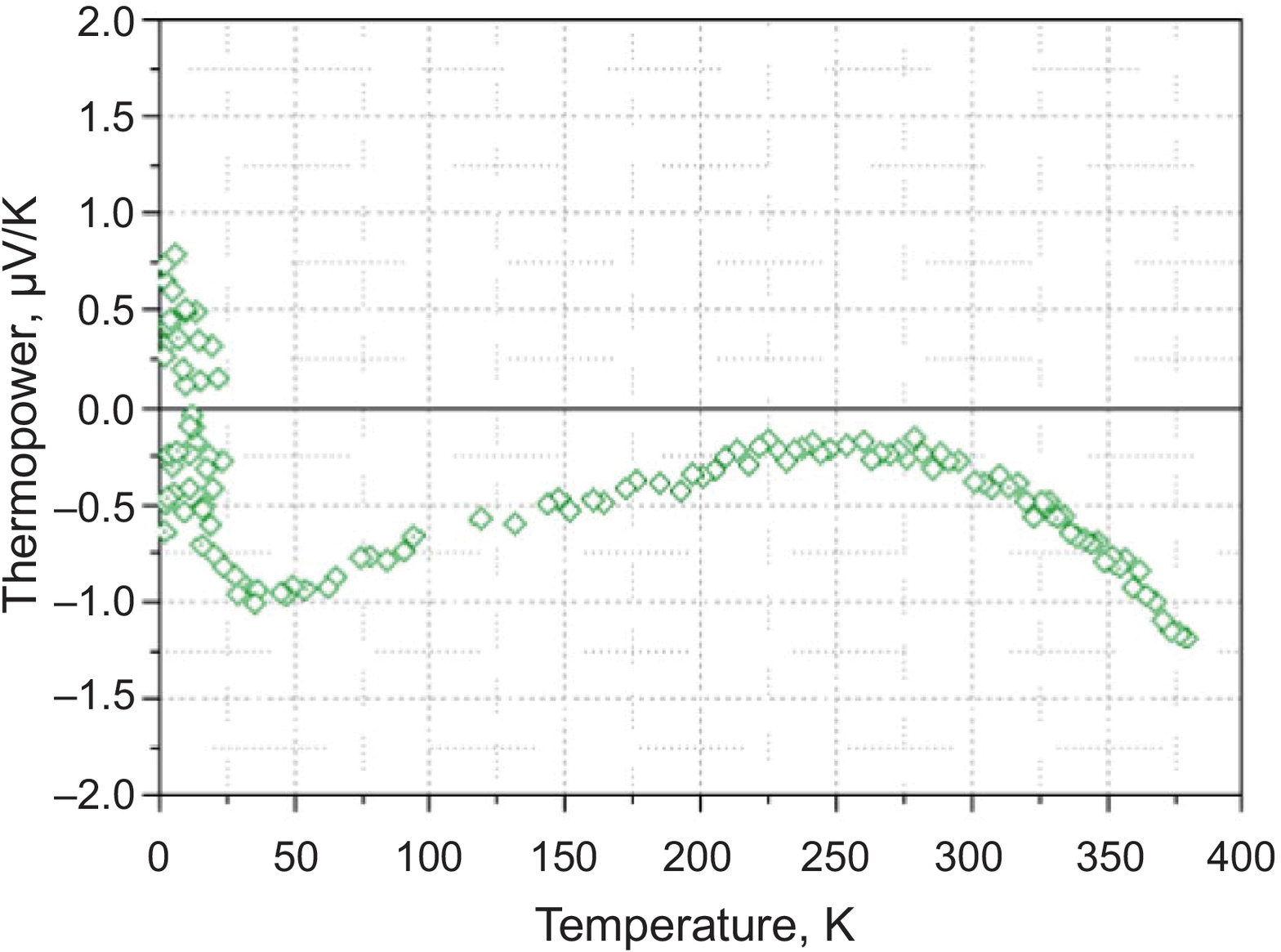}
\caption{The temperature dependence of  thermoelectric power  }
\label{seebeck}
\end{figure}

At increasing temperature, resistivity $\rho$ shows a power-law increase from the residual value $\rho_0$ = 1.3 $\mu \Omega\cdot$m (see Fig. 3).
At not too low temperatures it deviates from the $T^2$ behavior, so that we have $\rho-\rho_0 \sim T^{1.2}$, which indicates non-Fermi liquid features. At the same time, we did not found negative magnetoresistance characteristic for incoherent Kondo scattering (see, e.g., the results of Ref.\cite{Yatskar} for YbNi$_2$B$_2$C) in the field of 9 T starting from low $T$ up to room temperatures.


Thermal conductivity $\kappa$ (Fig. 4) increases linearly at low temperatures, then grows slowly and sharply increases again. The room temperature values of $\kappa$ of order of 10 W/m$\cdot$K as well as further  increase with temperature are usual for Yb- and Ce-based systems \cite{YbAgGe131,YbAgGe13}.

Thermoelectric power $S$ has a non-monotonous behavior and changes its sign at low temperatures (Fig. 5). The room temperature values of order of 1 $\mu V/K$ are typical for Yb-based systems like YbCu$_{2}$Sb$_{2}$, but are small in comparison with anomalous Ce systems (cf. Ref. \cite{YbAgGe131}). At the same time, much larger absolute values of $S$ are observed in skutterudite systems Yb$_{3}$Co$_{4}$Ge$_{13}$ and Yb$_{3}$Co$_{4}$Sn$_{13}$ \cite{YbAgGe13}.



\section{Discussion}

In intermetallic compounds including Ce and Yb elements the hybridization between the 4f- and itinerant conduction-band electrons induces the instability of magnetic moments and charge configurations.
In a number of Yb-based compounds, pressure-induced magnetic--nonmagnetic transitions were observed. The key point here is that Yb ions fluctuate between the nonmagnetic Yb$^{2+}$ ($J = 0$) and magnetic Yb$^{3+}$ ($J = 7/2$) states. Since the ionic volume for the Yb$^{3+}$ state is smaller than for Yb$^{2+}$, applying pressure or chemical pressure stabilizes the former configuration and induces the appearance of a magnetically ordered state, in contrast to Ce systems.

In this connection, our results on the system YbPd$_2 $Ge$_{2-x}$ can be treated in terms of the chemical pressure which changes with changing the germanium content.
In particular, unlike ferromagnetic compound  YbPd$_2$Ge \cite{NikiforovFTT}, the system YbPd$_2 $Ge$_{1.7}$ shows a nonmagnetic ground state, as well as YbPd$_2$Ge$_2$.
Thus the system passes a magnetic instability (quantum critical point), the ground state being determined by the competition between the magnetic RKKY interaction and the Kondo screening of magnetic moments. The data on specific heat of YbPd$_2 $Ge$_{1.7}$ demonstrate a moderate heavy-fermion behavior, although the factor $\gamma \simeq $ 100 mJ/mol K$^2$ is somewhat smaller in comparison with other anomalous Yb systems (see the Introduction).

The deviation from the $T^2$ Fermi-liquid behavior of resistivity seems to be more pronounced as compared to  YbPd$_2$Ge$_2$, but  weaker than in YbPd$_2$Ge \cite{NikiforovFTT}. The unusual behavior of  resistivity can be ascribed to closeness to the quantum critical point.
Sign change in the the thermoelectric power at low temperatures can be related to Kondo features (development of a coherent state) \cite{367}.


The authors are grateful to Yu.A. Koksharov for useful discussions and to A.V. Zarubin for the help in preparing the manuscript. The research was carried out within the state assignment of FASO of Russia (theme ``Quantum'' No. AAAA-A18-118020190095-4) and supported in part by Ural Division of RAS (project no. 18-2-2-11).

\end{document}